\documentclass[10pt]{article}
\usepackage[OE]{express}

\begin{document}
\title{Surface modeling for optical fabrication with linear ion source}

\author{Lixiang Wu,\authormark{*} Chaoyang Wei,\authormark{**} and Jianda Shao}

\address{Key Laboratory of Materials for High Power Laser, Shanghai Institute of Optics and Fine Mechanics, Chinese Academy of Sciences, Shanghai 201800, China}

\email{\authormark{*}wulixiang@siom.ac.cn} %% email address is required
\email{\authormark{**}siomwei@siom.ac.cn}
% \homepage{http:...} %% author's URL, if desired

%%%%%%%%%%%%%%%%%%% abstract and OCIS codes %%%%%%%%%%%%%%%%
%% [use \begin{abstract*}...\end{abstract*} if exempt from copyright]

\begin{abstract}
We present a concept of surface decomposition extended from double Fourier series to nonnegative sinusoidal wave surfaces, on the basis of which linear ion sources apply to the ultra-precision fabrication of complex surfaces and diffractive optics.
It is the first time that we have a surface descriptor for building a relationship between the fabrication process of optical surfaces and the surface characterization based on PSD analysis, which akin to Zernike polynomials used for mapping the relationship between surface errors and Seidel aberrations.
Also, we demonstrate that the one-dimensional scanning of linear ion source is applicable to the removal of surface errors caused by small-tool polishing in raster scan mode as well as the fabrication of beam sampling grating of high diffractive uniformity without a post-processing procedure.
The simulation results show that, in theory, optical fabrication with linear ion source is feasible and even of higher output efficiency compared with the conventional approach.
\end{abstract}

 % REPLACE WITH CORRECT OCIS CODES FOR YOUR ARTICLE, MINIMUM OF TWO; Avoid using the OCIS codes for “General” or “General science” whenever possible.
%For a complete list of OCIS codes, visit: https://www.osapublishing.org/oe/submit/ocis/

%%%%%%%%%%%%%%%%%%%%%%% References %%%%%%%%%%%%%%%%%%%%%%%%%

%%%%%%%%%%%%%%%%%%%%%%%%%%  body  %%%%%%%%%%%%%%%%%%%%%%%%%%

\section{Introduction}
The optics community has taken a lot of effort into the development of surface modeling methods for optical design~\cite{forbes2007shape,jester2011b,jester2012wavelet,forbes2013fitting,ferreira2016orthogonal}.
Unfortunately, none of those methods are developed to improve manufacturability~\cite{forbes2011manufacturability} or to reduce the difficulty of fabrication of optical surfaces.
For example, Zernike polynomials are widely applied to describe surface errors and express wavefront data in the field of optical fabrication and testing.
Because there is a theoretical relationship between Zernike coefficients and Seidel aberrations often observed in optical tests~\cite{tyson1982conversion}.
However, basically, the standard 36-term Zernike polynomial set does not aim to or help to optimize the process of optical fabrication or polishing.

When it comes to ultra-precision fabrication of optical surfaces, ion beam technologies play a key role in improving surface precision to extreme.
As one of the most precise methods, ion beam figuring (IBF)~\cite{xie2015ion} is employed for finishing lithography optics and telescope mirrors.
IBF is realized with a dwell time algorithm akin to that used in computer controlled optical surfacing (CCOS).
Longer time the ion beam dwelling at a point results in more material removal around the point, thus it allows correction of surface figures through controlled variations of scanning velocity.
In particular, one-dimensional IBF~\cite{zhou2016one}, or sometimes called ion beam profiling~\cite{peverini2010ion}, applies to the fabrication of elongated synchrotron optics such as aspherical X-ray mirrors.
Compared with the conventional IBF, one-dimensional IBF is generally implemented by a simpler algorithm meanwhile the long-rectangular-shaped ion beam adopted in one-dimensional case leads to higher output efficiency under the condition for ensuring similar finishing precision.

Inspired by Zernike polynomials bridging the relationship between surface errors and optical aberrations, we propose a surface modeling method on the basis of double Fourier series~\cite{moricz1989convergence} to narrow the gap between optical fabrication and surface characterization.
The Fourier series decomposition of an optical surface produces a set of wave surfaces with a sinusoidal profile, which by and large are of different periods, amplitudes, and propagation directions.
Conceptually, we can sequentially fabricate the decomposed wave surfaces by linear scanning with a linear ion source and the superposition of those fabricated wave surfaces finally build up a surface that approximates to the desired optical surface.
Moreover, the power spectral density (PSD) analysis is the statistical analysis of the spatial-distributed wave surfaces described as components of double Fourier series.
As to the characterization of optical surfaces, the PSD function is typically utilized~\cite{youngworth2005overview}, especially for evaluating errors in the mid-spatial frequency (MSF) range.
So, we realized, a surface descriptor based on Fourier series can build a relationship between the technique of optical fabrication with linear ion source and the PSD-based surface characterization method.

This paper introduces the basic theory of surface modeling and decomposition, then illustrates the principle of optical fabrication with linear ion source, and finally demonstrates two applications, i.e., fabrication of large-aperture beam sampling gratings (BSGs) and removal of errors in the MSF range.

\section{Surface decomposition}

A continuous smoothing surface $f(x,y)$ can be described as
\begin{equation}
f(x,y) = \sum\nolimits_{m \in \mathbb{Z}}\sum\nolimits_{n \in \mathbb{Z}} e^{i\omega_1mx}C_{m,n}e^{i\omega_2ny},
\end{equation}
where $f(x,y)$ is periodic by $T_1$ in X direction and $T_2$ in Y direction; $\omega_1 = 2\pi/T_1$, $\omega_2 = 2\pi/T_2$; $C_{mn}$ is a complex number.
In particular, $f(x,y)$ is a real function that is conjugate symmetric thus the imaginary part can be omitted.
After adding a proper piston to every component, $e^{i\omega_1mx}C_{m,n}e^{i\omega_2ny}$, we get a slightly elevated surface, $S(x,y)$, that is suitable for indicating the spatial distribution of material removals and is displayed as
\begin{equation} \label{eq:removal_surface1}
S(x,y) = \sum\nolimits_{m \in \mathbb{Z}}\sum\nolimits_{n \in \mathbb{Z}}(e^{i\omega_1mx}C_{m,n}e^{i\omega_2ny} + \vert C_{m,n} \vert),
\end{equation}
where $\vert C_{m,n} \vert$ represents the piston or extra material removal; note that the arbitrary component $P_{m,n}(x,y) = e^{i\omega_1mx}C_{m,n}e^{i\omega_2ny} + \vert C_{m,n} \vert \ge 0$.
Since ion etching is a subtractive process rather than an additive process, the desired material removal should be nonnegative thus it requires that $P_{m,n}(x,y) \ge 0$.
Continued by Equation~\ref{eq:removal_surface1}, we put forward a new formula to decrease the number of scan strokes and improve the reachability of optical surfacing, which is represented as
\begin{equation} \label{eq:removal_surface2}
S'(x,y) = \sum\nolimits_{m \in \mathbb{Z}_{\geq 0}}\sum\nolimits_{n \in \mathbb{Z}} (2 P_{m,n} + \varepsilon),
\end{equation}
where $m$ and $n$ not equal to $0$ simultaneously; $\varepsilon$ denotes a small positive quantity and represents the compensated material removal.
Since there is an optimum removal in ion-beam figuring~\cite{zhou2010optimum}, $\varepsilon$ usually is not infinitesimal in order to make sure that the scan speed is properly limited.
The component $P^{'}_{m,n} = 2 P_{m,n} + \varepsilon$, indicating the decomposed distribution of material removals, is shown as a sinusoidal wave surface, where the propagation direction is $(\frac{m}{T_1}, \frac{n}{T_2})$ or $(-\frac{m}{T_1}, -\frac{n}{T_2})$ and the period is $(\frac{m^2}{T_1^2} + \frac{n^2}{T_2^2})^{-\frac{1}{2}}$.

\subsection{Calculation of Fourier coefficients via Fast Fourier Transform}

\begin{equation} \label{eq:coeff}
C_{M \times N} = \frac{\mathrm{FFT2}(S_{M \times N})}{MN},
\end{equation}
where $S_{M \times N}$ represents a material removal map by M times N, which is the discrete representation of $S^{'}(x,y)$; two-dimensional Fast Fourier Transform (denoted as FFT2 in this paper) algorithm is used for calculating the matrix of Fourier coefficients, $C_{M \times N}$.
To exclude the DC component, $P_{0,0}$, it needs to assign that: $C_{M \times N}(1,1) = 0$.

\subsection{Error analysis based on PSD function}
The material removal map can be effectively approximated by the symmetric rectangular partial sum of double Fourier series~\cite{moricz1989convergence}.
The ($m_1, n_1$)-th symmetric rectangular partial sum is defined as
\begin{equation}
S'_{m_1n_1} = S'(x,y; m, n), \qquad 0 \le m \le m_1, 0 \le \vert n \vert \le n_1,
\end{equation}
then we have the truncated errors
\begin{equation}
E_{m_1n_1} = S'(x,y) - S'_{m_1n_1}.
\end{equation}

To evaluate the feasibility of the approximation, we use PSD function to analyze the truncated errors referring to as surface errors.
The PSD analysis of the truncated errors can be implemented by the two-dimensional FFT algorithm, which is displayed as follows:
\begin{equation} \label{eq:PSDofErrors}
\mathrm{PSD}(E_{m_1 n_1}) = \frac{T_1 T_2}{M^2 N^2} \mathrm{FFT2}(E_{m_1 n_1}).
\end{equation}

\section{Optical fabrication with linear ion source}

Reactive ion etching (RIE) is a useful method in the fabrication of diffraction optics.
Recently we fabricated large-aperture BSGs by the RIE process, where the ion beam emitted from linear ion source scans over BSGs along a linear guide meanwhile the rotary leaf scans along the elongated footprint of ion beam, thus the spatial distribution of etch depths is finely adjusted for improving the diffractive uniformity of BSG~\cite{wu2015algorithms,wu2016fine}.
Now we try to exploit more applications by combining optical fabrication technique with linear ion source and the aforementioned surface modeling method.

\begin{figure}[htbp]
\centering\includegraphics[width=.65\columnwidth]{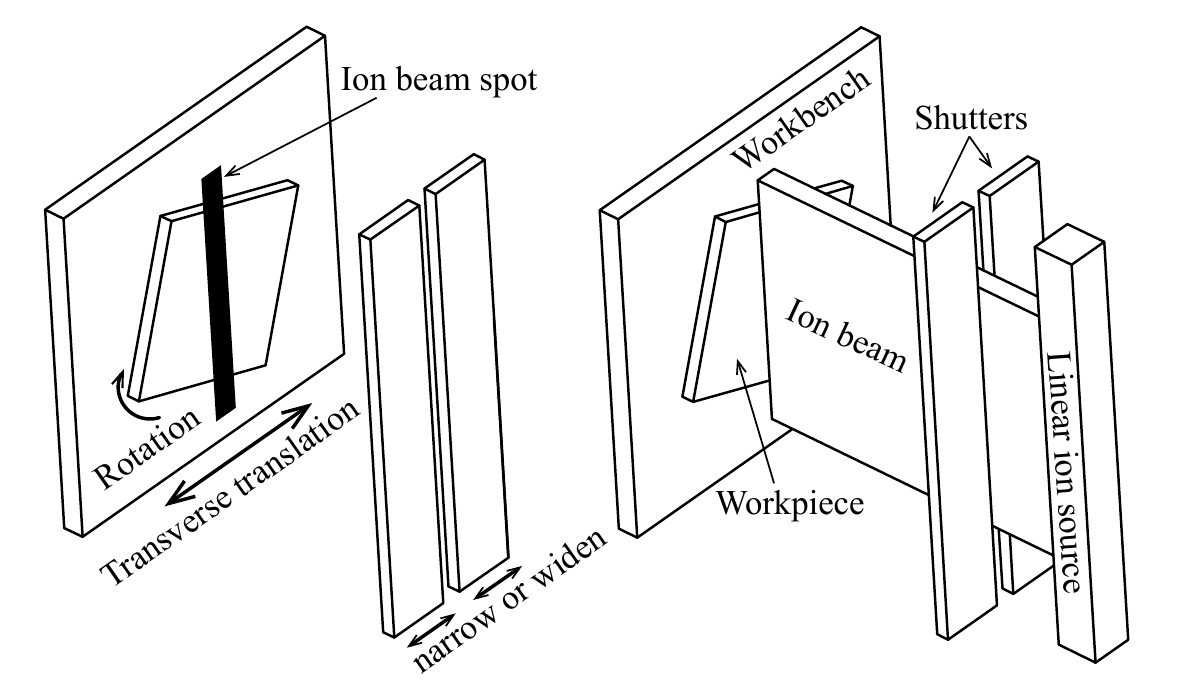}
\caption{Schematic of optical fabrication with linear ion source.}
\label{fig:optsurf}
\end{figure}

Fig.~\ref{fig:optsurf} illustrates the basic principle of optical fabrication with linear ion source and also serves as a schematic of the fabrication of BSGs.
The two shutters can collimate the ion beam emitted from the linear ion source and adjust the width of ion beam spot (see Fig.~\ref{fig:optsurf}) projected on the workpiece or BSG substrate.
The workpiece Mounted on a workbench spins about its center normal line and moves in the transverse horizontal direction in accompany with the workbench.
That means the ion beam having an elongated rectangular footprint can scan over the workpiece along the arbitrary direction on the workpiece surface.

Linear ion sources, or ion sources with a large-aperture rectangular grid, are suitable for ion figuring of optical surfaces with mainly a one-dimensional profile such as synchrotron reflective mirrors.
Fortunately, we only need to fabricate wave surfaces with a one-dimensional sinusoidal profile with the help of surface decomposition.
A set of decomposed wave surfaces linearly build up a superposed surface approximating to the desired surface.
Moreover, we have compared two approach in our previous work~\cite{wu2016variable} to calculate the dwell time distribution in corresponding with a sinusoidal profile.

\section{Applications}

\subsection{Fabrication of large-aperture BSGs}

\begin{figure}[htbp]
\centering\includegraphics[width=.75\columnwidth]{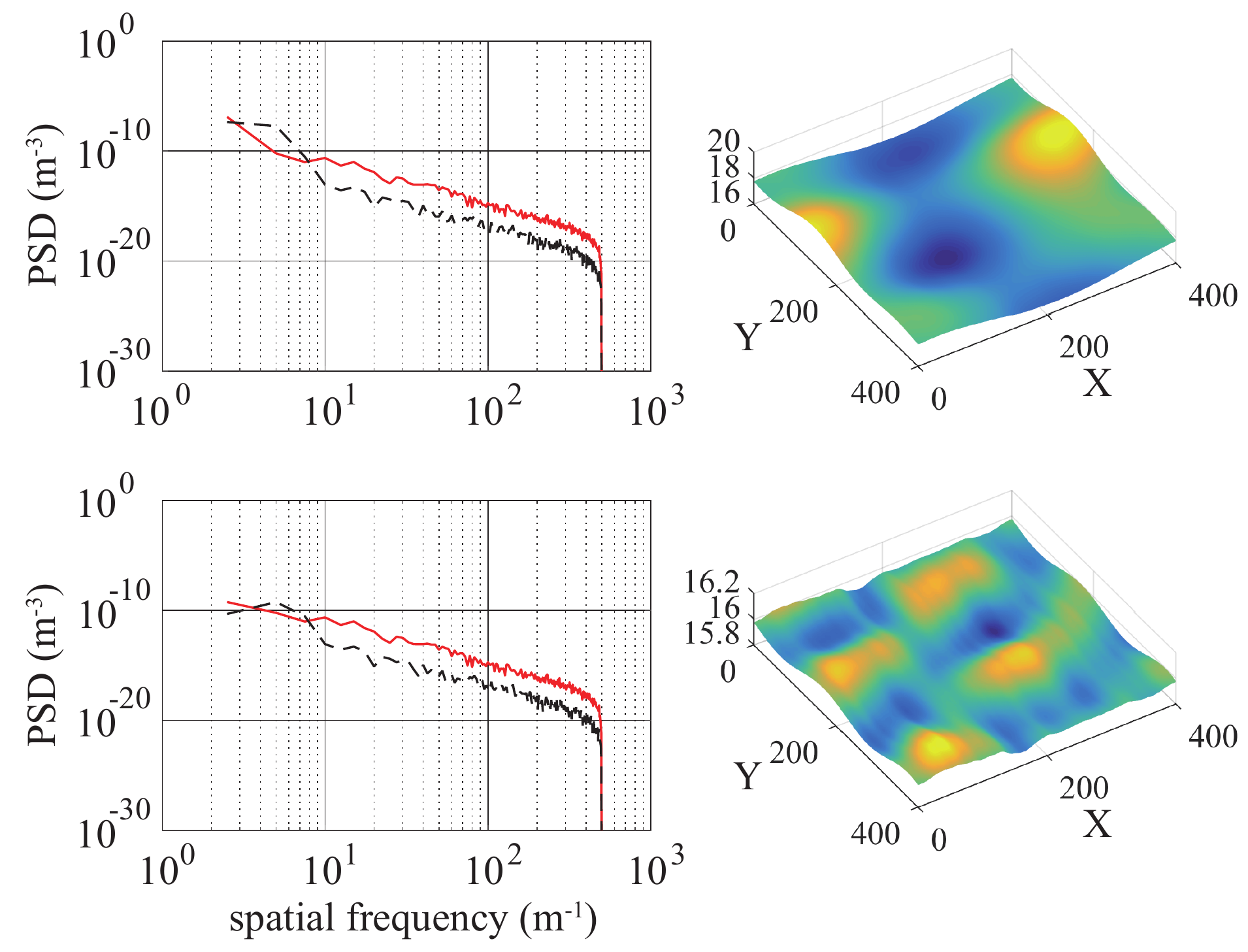}
\caption{Surface decomposition based on PSD analysis. The upper shows the original etch depth map and its PSD plots along X direction (solid curve) and Y direction (dotted curve); the lower shows the etch depth residuals obtained by removing $S_1$ (see Fig.~\ref{fig:decomp}) from the etch depth map.}
\label{fig:reduce}
\end{figure}

\begin{figure}[htbp]
\centering\includegraphics[width=.7\columnwidth]{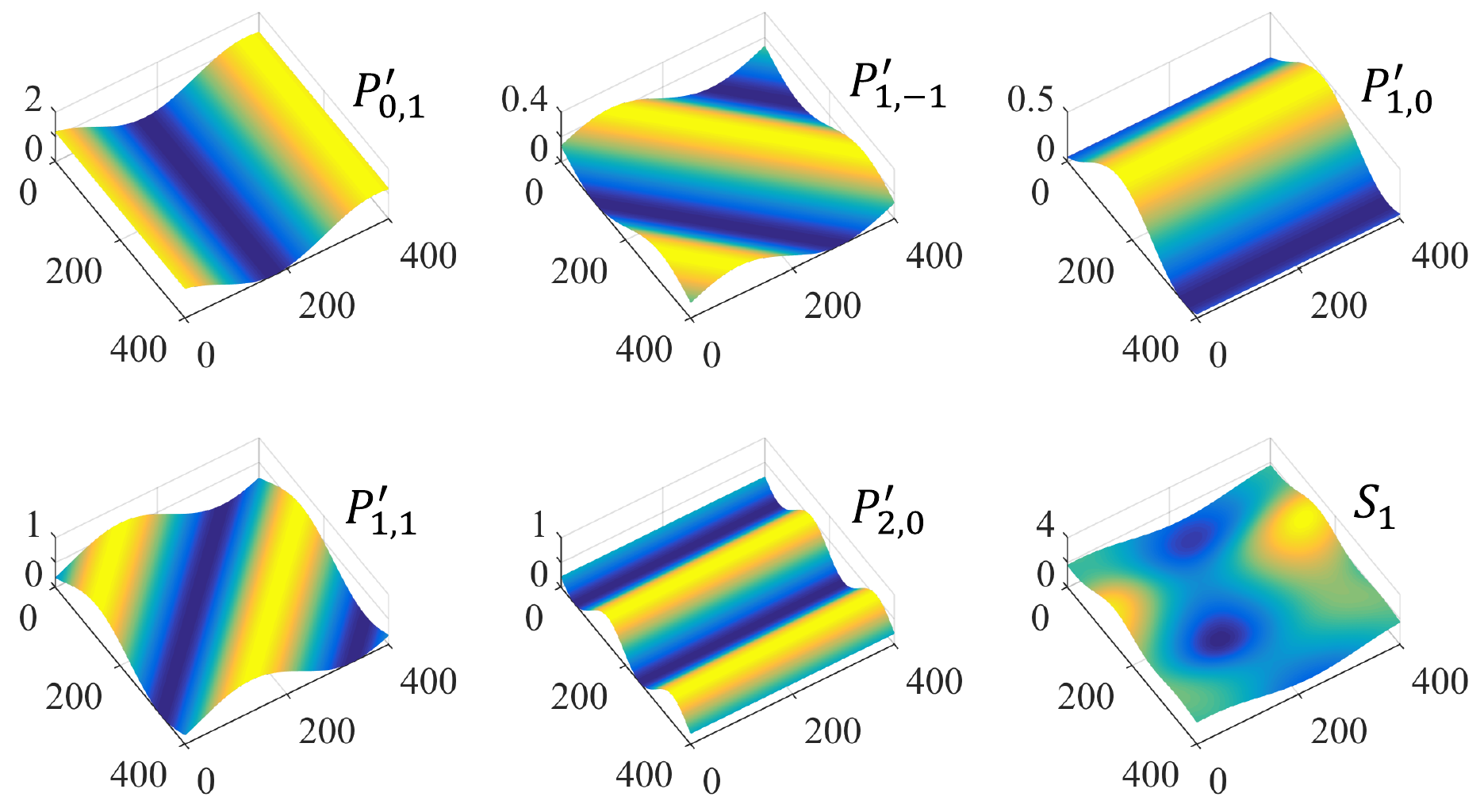}
\caption{Sinusoidal wave surfaces ($P^{'}_{0,1}, P^{'}_{1,-1}, P^{'}_{1,0}, P^{'}_{1,1}, P^{'}_{2,0}$) and their summation ($S_{1}$).}
\label{fig:decomp}
\end{figure}

In the fabrication of large-aperture BSGs, the spatial distribution of etch depths should be finely adjusted to improve the diffractive uniformity of BSG.
The mesh graph on the top right of Fig.~\ref{fig:reduce} is a typical etch depth map, which describes the desired etch depths spatially distributed on the BSG substrate.
The tolerance analysis indicates that the RMS of etch depth differences should be within 0.4~nm, which means the RMS of truncated errors, $E_{m_1, n_1}$, is not greater than 0.4~nm.
As is shown on the top left of Fig.~\ref{fig:reduce}, the PSD in the low-spatial frequency range is significantly higher than that in the mid/high-spatial frequency range, which indicates that the first several components or decomposed wave surfaces mainly contribute to the surface form.
After PSD analysis, as is shown in Fig.~\ref{fig:decomp}, the first 5 sinusoidal wave surfaces including $P^{'}_{0,1}, P^{'}_{1,-1}, P^{'}_{1,0}, P^{'}_{1,1}, P^{'}_{2,0}$ are selected and the summation ($S_1$) of the 5 components is well approximated to the desired etch depth map.
The etch depth differences, obtained by removing $S_1$ from the desired etch depth map, is shown on the lower part of Fig.~\ref{fig:reduce}, where the peak-to-valley value is 0.19~nm and the average of etch depth differences is 15.95~nm.

For this example, to fabricate a BSG of high diffractive uniformity, it needs to carry out at least 6 times of linear scanning of ion beam.
To start with removing a uniform layer of thickness of 15.95~nm, we then fabricate the 5 wave surfaces sequentially by linear scanning of ion beam.
As is shown in Fig.~\ref{fig:decomp}, for the 5 sinusoidal wave surfaces, the rotary directions are (0, 1), (1, -1), (1, 0), (1, 1), and (2, 0); the periods are 400~mm, 283~mm, 400~mm, 283~mm, and 100~mm.
In the process of ion beam etching, the BSG should be rotated according to the rotary directions and the width of ion beam spot should be less than a half of the period of wave surface.

As is demonstrated above, optical fabrication with linear ion source can be used to finely adjust etch depths of large-aperture diffractive optics based on the concept of surface decomposition.

\subsection{Removal of errors in the MSF range}

\begin{figure}[htbp]
\centering\includegraphics[width=.75\columnwidth]{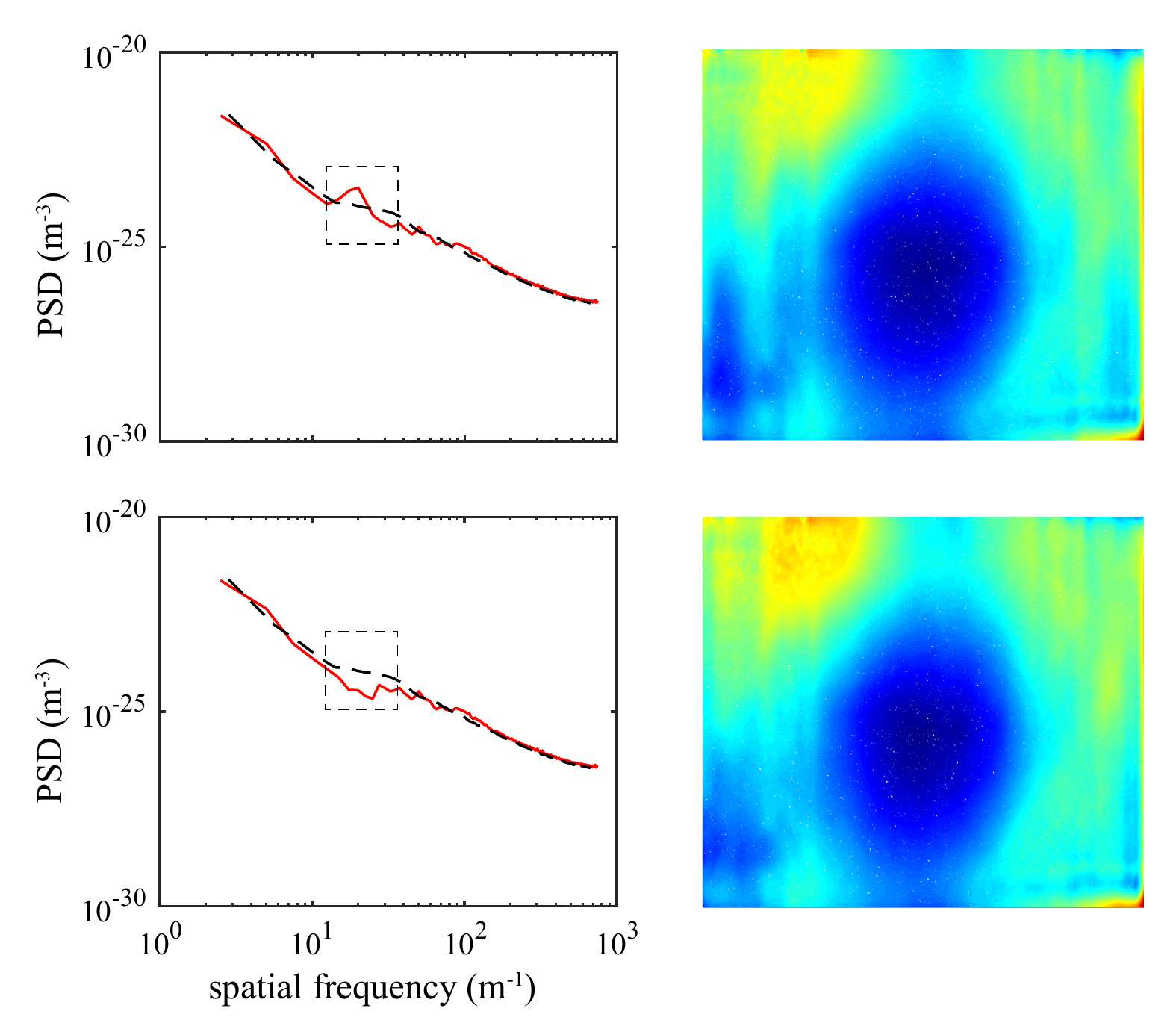}
\caption{Removal of errors in a specific range of spatial frequencies. The right half part shows the original surface map and the smoothed surface map; the left half part shows the corresponding PSD curves, where the solid curve and the dotted curve represent the PSD functions calculated along the horizontal and the vertical, respectively.}
\label{fig:errors}
\end{figure}

We prepared a large-aperture optical surface ($525px\times594px, 0.673mm/px$) polished by CCOS with small tools.
The measured surface map is shown on the upper right of Fig.~\ref{fig:errors}.
And more MSF errors are found in the horizontal direction than those in the vertical direction, which is attributed to the raster scan path adopted in the small-tool polishing process.
The corresponding PSD plots are arranged on the left of Fig.~\ref{fig:errors}, where the solid curve and the dotted curve represent the PSD functions along the horizontal and the vertical, respectively.
As is shown on the upper left of Fig.~\ref{fig:errors}, there is a high peak at the solid PSD curve in the spatial frequency of around $20~m^{-1}$.
In general, a peak at the PSD curve indicates more errors existing in the spatial frequency range around the peak.
To remove errors in a specific range of spatial frequencies, we should target the errors displayed as sinusoidal wave surfaces and then remove them one by one via optical surfacing with linear ion source. 

\begin{figure}[htbp]
\centering\includegraphics[width=.75\columnwidth]{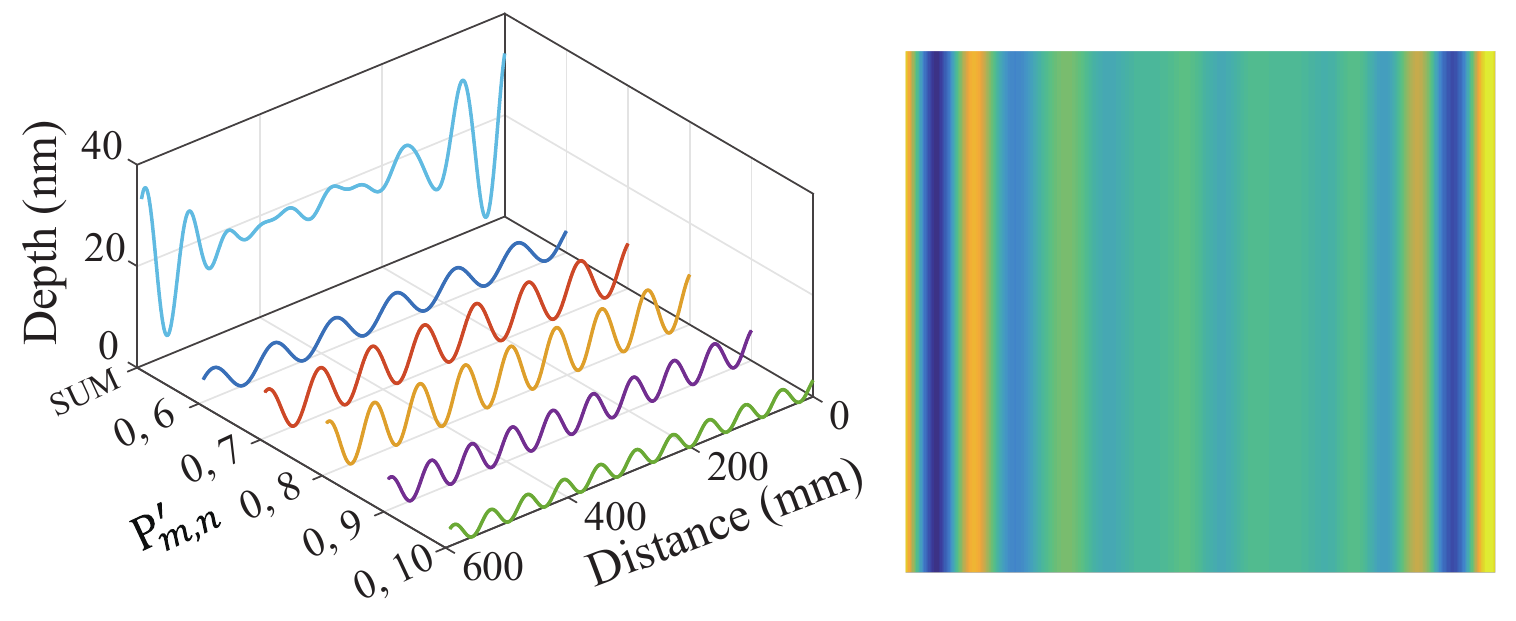}
\caption{A set of sinusoidal wave surfaces, $\{P'_{0,n} | n=6,7,8,9,10\}$, summing up to remove errors in the spatial frequency range around $20~m^{-1}$.}
\label{fig:waves}
\end{figure}

Given that $T_1 = 525px \times 0.673mm/px$; $T_2 = 594px \times 0.673mm/px$;
$(\frac{m^2}{T_1^2} + \frac{n^2}{T_2^2})^{-\frac{1}{2}} = \frac{1}{20~m^{-1}},$
and provided that $m = 0$ (only the errors distributed along the horizontal are of interest), then we have $n = 8$, which refers to $P'_{0,8}$.
Actually, to remove the peak totally, it requires a set of sinusoidal wave surfaces.
Fig.~\ref{fig:waves} shows a set of sinusoidal wave surfaces, $\{P'_{0,n} | n=6,7,8,9,10\}$, stacking up to remove errors in the spatial frequency range around $20~m^{-1}$.
The summation, $\sum_{n=6}^{10}{P'_{0,n}}$, represents the to-be-etched material removal, which is shown on the right of Fig.~\ref{fig:waves}.
After reducing the to-be-etched material removal from the original surface map, we obtain a surface map with less errors in the frequency range nearby $20~m^{-1}$ (shown on the bottom right of Fig.~\ref{fig:errors}) and the peak disappears in the corresponding PSD plot.

The simulation result indicates that, theoretically, one-dimensional polishing is able to get rid of MSF errors spatially distributed vertical to the scanning direction.
It also suggests that ultra-precision optical surfacing with linear ion source is a feasible and preferred approach to reduce errors in MSF range.
Because optical surfacing with linear ion source is a superposition of multiple one-dimensional polishing processes and there is no discrete stitching of polishing spots in the course of polishing~\cite{li2009stitching}.
As illustrated in Fig.~\ref{fig:optsurf}, the aperture of optical element is less than the length of rectangular beam spot thus the ion beam can fully cover and scan over the optical surface without stitching along the length direction.

\section{Conclusion}

We propose a surface modeling method based on double Fourier series and the concept of surface decomposition makes it feasible to fabricate an ultraprecise optical surface with linear ion source.
Also, we demonstrate two applications.
The first application shows that optical fabrication with linear ion source can be applied to direct fabricating a BSG of high diffractive uniformity without post-processing (for example, chemical mechanical polishing) for correcting the spatial distribution of diffraction efficiency.
The second application presents a new approach to remove surface errors in MSF range and we think the basic concept can be realized with other advanced polishing tools.
Moreover, optical fabrication with linear ion source will significantly improve the working efficiency compared with the conventional IBF, though small round beam has a higher reachability in the fabrication of complex surfaces.

\section*{Funding}

Youth Innovation Promotion Association of the Chinese Academy of Sciences.

\section*{Acknowledgments}

We thank Mr. Zhentong Liu for the help in drawing the schematic of optical fabrication with linear ion source.

\end{document}